\documentstyle[preprint,aps,epsfig,psfig]{revtex}
 
\begin{document}

\draft

\title{Analysis of the Decays $B\rightarrow \pi \pi$ and $\pi K$
with QCD Factorization in the Heavy Quark Limit
\footnote{Supported in part by National Natural Science
Foundation of China and State Commission of Science and 
Technology of China}}
\vspace{2cm}

\author{
Dongsheng Du${}^{1,2}$, Deshan Yang${}^{2}$  and Guohuai Zhu${}^{2}$
\footnote{Email: duds@hptc5.ihep.ac.cn, yangds@hptc5.ihep.ac.cn,
zhugh@hptc5.ihep.ac.cn} }
\address{${}^1$ CCAST (World Laboratory), P.O.Box 8730, Beijing
100080, China\\
${}^2$ Institute of High Energy Physics, Chinese Academy of Sciences,
 P.O.Box 918(4), Beijing 100039, China
 \footnote{Mailing address}}

\date{\today}

\maketitle

\begin{abstract}
\indent

  In this work, we calculate the branching ratios and CP asymmetries of
the decays of $B \rightarrow \pi \pi$ and $\pi K$ with the frame of QCD
factorization in the heavy quark limit. We also compare the results with the
estimates by using generalized factorization and experimental measurements.

\end{abstract} 

\vspace{1.5cm}
 
{\bf PACS numbers 13.25.Hw 12.38.Bx}

\newpage

\narrowtext
\tighten

In the past two years, the CLEO collaborations\cite{CLEO}  had improved 
their measurements for the branching ratios of $B \rightarrow \pi \pi$ 
and $\pi K$ for several times. The latest results of branching ratios of 
these modes are:
\begin{eqnarray}
Br(B^{\pm} \rightarrow \pi^{0} \pi^{\pm}) &<& 12.7 \times 10^{-6}
~,\nonumber\\
Br(B^0 \rightarrow \pi^{+} \pi^{-}) &=& (4.3^{+1.6}_{-1.4} \pm 0.5)
\times 10^{-6} ~,\nonumber \\
Br(B^{\pm} \rightarrow K^{\pm} \pi^{0})&=&(11.6^{+3.0+1.4}_{-2.7-1.3})
\times 10^{-6} ~,\nonumber \\
Br(B^{\pm} \rightarrow K^{0} \pi^{\pm})&=&(18.2^{+4.6}_{-4.0}\pm 1.6) 
\times 10^{-6}  ~,\nonumber \\
Br(B^0 \rightarrow K^{\pm} \pi^{\mp})&=&(17.2^{+2.5}_{-2.4}\pm 1.2) \times
10^{-6} ~,\nonumber \\
Br(B^0 \rightarrow K^{0} \pi^{0})&=&(14.6^{+5.9+2.4}_{-5.6-3.3}) \times 
10^{-6}~.
\end{eqnarray}
These two-body non-leptonic charmless B decay modes play a very 
important role in studying CP violation and the heavy flavor physics. 
Theoretically, due to our ignorance of hadronization, it is difficult
to relate CP violations with the parameters of fundamental theory.
Effective Hamiltonian is our basic tool, but we do not know how to
calculate hadronic matrix element from first principles.
Conventionally we resort to naive factorization assumption\cite{BSW},
which expresses the hadronic matrix element in terms of various meson decay
constants and form factors. However, it is well known that non-factorizable
contribution really exists and can not be neglected numerically. To
remedy factorization hypothesis, Ref.\cite{chy,Ali} introduced a 
phenomenological parameter $N_{eff}$, which is commonly called generalized
factorization. Because in principle $N_{eff}$ is process-dependent, it is 
still not a satisfactory approach.  

In last year, Beneke {\it et al.} \cite{Beneke} gave a NLO calculation
of $B \rightarrow \pi \pi$ in heavy quark limit. In this limit, the 
hadronic matrix elements for $B \rightarrow \pi \pi$
can expanded by the powers of $\alpha_s$ and $\Lambda_{QCD}/m_b$ as follows:

\begin{equation}
\langle \pi \pi \vert Q \vert B \rangle = \langle \pi \vert j_1 \vert
\vert B\rangle \langle \pi \vert j_2 \vert 0\rangle \cdot
[1+\Sigma r_n \alpha_s^n + {\cal{O}}(\Lambda_{QCD}/m_b)],
\end{equation}
where $Q$ is a local four quark operator in the weak effective 
Hamiltonian and $j_{1,2}$ are bilinear quark currents. Neglecting
the power contribution of $\Lambda_{QCD}/m_b$, they pointed
out that in the heavy quark limit the radiative corrections at the order
of $\alpha_s$ can be calculable with PQCD method. Furthermore, 
the 'non-factorizable' contributions from hard scattering with 
spectator quark in B meson can also be calculable within the frame of PQCD. 
Then the decay amplitude can be expressed by the convolutions of the 
hard-scattering kernels and several light-cone distribution amplitudes
of the mesons. So all of these can be summarized into a factorization
formula for $B \rightarrow \pi \pi$ as follows:

\begin{equation}
\langle \pi(p^{\prime}) \pi(q) \vert Q_i \vert B(p) \rangle =
F^{B \rightarrow \pi}(q^2) \int \limits_0^1 dx T^I_i(x) \Phi_{\pi}(x)
+\int \limits_0^1 d\xi dx dy T^{II}(\xi,x,y) \Phi_B(\xi) \Phi_{\pi}(x)
\Phi_{\pi}(y).
\end{equation}
We call this factorization formalism as QCD factorization. 
In the above formula, $\Phi_B(\xi)$ and $\Phi_{\pi}(x)$ are the 
leading-twist light-cone distribution amplitudes of B and pion mesons 
respectly, and the $T^{I,II}_i$ denote hard-scattering kernels which are 
calculable in perturbative theory. Beneke {\it et al.} regarded 
the spectator quark as a soft quark translated to one pion in final state
unless it undergoes a hard interaction,
so the transition form factor $F^{B \rightarrow \pi}(q^2)$ can not be 
calculated in the perturbative frame, and it will be survived as a free 
nonperturbative parameter in QCD factorization. $T^I_i$ at the order of 
$\alpha_s$ includes two topological classes, one is hard gluon scattering 
like vertex corrections which is described by the first four diagrams in 
Fig.1, and we call it as hard-scattering corrections below; the other is 
called penguin correction, which is described by diag.(e) and (f) in
Fig.1.  $T^{II}_i$ denotes the hard spectator scattering contribution 
which is described by the last two diagrams in Fig.1. In this frame,
they neglected W-annihilation and W-exchange topologies, so they
show that 'non-factorizable' contributions in the conventional
factorization are calculable in heavy quark limit. Then we do
not need to employ a phenomenological parameter $N_{eff}$ to compensate
the 'non-factorizable' effects in principle.  Thus we will apply this
approach to calculate the branching ratios and CP asymmetries for the decays
$B \rightarrow \pi \pi$ and $\pi K$ in this paper.
Furthermore, since the effects of electroweak penguins can not be
neglected in some decay modes of $B \rightarrow \pi K$, we
will use full effective weak Hamiltonian including electroweak penguin
operators and add some one-loop QED penguin correction to the calculation
of hadronic matrix elements of effective operators which can also be
described by diag.(e) and diag.(f) (replacing gluon with photon) in Fig.1 
in our computation.
      
The $\vert\Delta B\vert=1$ effective Hamiltonian is given by\cite{buras} 
\begin{equation}
{\cal{H}}_{eff}= \frac{G_F}{\sqrt{2}}
 \left[ \sum_{q=u,c} v_q \left(  C_1(\mu) Q^q_1(\mu)+ C_2(\mu)Q^q_2(\mu)
  + \sum_{k=3}^{10} C_k(\mu)Q_k(\mu)  \right) \right]+h.c.,
\end{equation}
where
$v_q=V_{qb}V_{qd}^{*}$(for $b\rightarrow d$ transition) or
$v_q=V_{qb}V_{qs}^{*}$(for $b\rightarrow s$ transition)
and $C_i(\mu)$ are Wilson coefficients which have been evaluated to 
next-to-leading order approximation.
 In the Eq.(1), the four-quark operators $Q_i$ are given by
\begin{equation}
\begin{array}{l}
\begin{array}{ll}
Q^u_1= ( \bar{u}_{\alpha} b_{\beta} )_{V-A}
         ( \bar{q}_{\beta} u_{\alpha} )_{V-A}&
Q^c_1= ( \bar{c}_{\alpha} b_{\beta} )_{V-A}
         ( \bar{q}_{\beta} c_{\alpha} )_{V-A}\\
Q^u_2= ( \bar{u}_{\alpha} b_{\alpha} )_{V-A}
         ( \bar{q}_{\beta} u_{\beta} )_{V-A}&
Q^c_2= ( \bar{c}_{\alpha} b_{\alpha} )_{V-A}
         ( \bar{q}_{\beta} c_{\beta} )_{V-A}\\
Q_3= (\bar{q}_{\alpha} b_{\alpha} )_{V-A}
      \sum\limits_{q'}
     ( \bar{q}^{'}_{\beta} q^{'}_{\beta} )_{V-A}&
Q_4= (\bar{q}_{\beta} b_{\alpha} )_{V-A}
      \sum\limits_{q'}
     ( \bar{q}^{'}_{\alpha} q^{'}_{\beta} )_{V-A}\\
Q_5= (\bar{q}_{\alpha} b_{\alpha} )_{V-A}
      \sum\limits_{q'}
      ( \bar{q}^{'}_{\beta} q^{'}_{\beta} )_{V+A}&
Q_6= (\bar{q}_{\beta} b_{\alpha} )_{V-A}   
      \sum\limits_{q'}
     ( \bar{q}^{'}_{\alpha} q^{'}_{\beta} )_{V+A}\\
Q_7= \frac{3}{2} (\bar{q}_{\alpha} b_{\alpha} )_{V-A}
      \sum\limits_{q'} e_{q'}
     ( \bar{q}^{'}_{\beta} q^{'}_{\beta} )_{V+A}&
Q_8=\frac{3}{2}  (\bar{q}_{\beta} b_{\alpha} )_{V-A}
   \sum\limits_{q'} e_{q'}
    ( \bar{q}^{'}_{\alpha} q^{'}_{\beta} )_{V+A}\\
Q_9= \frac{3}{2} (\bar{q}_{\alpha} b_{\alpha} )_{V-A}
      \sum\limits_{q'} e_{q'}
    ( \bar{q}^{'}_{\beta} q^{'}_{\beta} )_{V-A}&
Q_{10}=\frac{3}{2}  (\bar{q}_{\beta} b_{\alpha} )_{V-A}
      \sum\limits_{q'} e_{q'}
     ( \bar{q}^{'}_{\alpha} q^{'}_{\beta})_{V-A}\\
\end{array} \\
      
\end{array}
\end{equation}
with $Q^q_1$ and $Q^q_2$ being the tree operators, $Q_3-Q_6$ the QCD
penguin operators and $Q_7-Q_{10}$ the electroweak penguin operators.
With the renormalization group method, we can evolve the Wilson coefficients
$C_i(\mu)$ from the scale $\mu=m_W$ to $\mu \approx m_B$. Because we will give
a NLO calculation here, $b\rightarrow s(d) g$ and $b \rightarrow s(d) \gamma$
effective Hamiltonian must be included. They are
\begin{equation}
{\cal{H}}_{eff}^{\prime} =
-\frac{G_F}{\sqrt{2}} v_t 
[C_{7\gamma} Q_{7\gamma} + 
C_{8G}Q_{8G}] + h.c., 
\end{equation}
where 
\begin{equation}
Q_{7\gamma}=\frac{e}{8\pi^2} m_b \bar{q}_{\alpha} \sigma^{\mu\nu} 
(1+\gamma_5) b_{\alpha} F_{\mu\nu}, ~~
Q_{8G}=\frac{g}{8\pi^2} m_b \bar{q}_{\alpha} \sigma^{\mu\nu} 
t^{a}_{\alpha \beta} b_{\beta} G^a_{\mu\nu}, ~~(q=d~ {\rm or} ~s).
\end{equation}  

Following the method of Ref.\cite{Beneke}, we try to evaluate
the matrix elements of $Q_i$ to the order of $\alpha_s$ and $\alpha_{em}$.
In fact, this work is to calculate hadronic matrix
elements of $Q_i$ to one loop. In quark level, Ref\cite{fleischer} gives an
expression

\begin{equation}
\langle \bf{Q} \rangle =
[\hat{\bf{1}}+\frac{\alpha_s}{4\pi} \hat{m}_s
+\frac{\alpha_{em}}{4\pi} \hat{m}_{e}]\cdot \langle \bf{Q} \rangle_0.
\end{equation}
Here $\hat{m}_s$ and $\hat{m}_e$ represent the one loop corrections of 
QCD and QED respectly. These corrections are divided into two classes.
One is hard-scattering correction, the other is penguin correction. While
the external quarks are on mass-shells, the hard-scattering corrections
bring infrared divergences which can not vanish after summing over all
this kind of perturbative diagrams no matter with gluon or photon exchange.
However, in hadron level, with QCD factorization in the heavy quark
limit, these infrared divergences from hard-scattering exchanging with
gluon can be canceled after summing over all this kind of perturbative
diagrams (the first four diagrams in Fig.1). This has been shown in
Ref.\cite{Beneke}. But for the case of QED hard-scattering 
(exchanging photon), the infrared divergences can not be canceled
after summing over all diagrams even in the heavy quark limit. It is
because that the color structure of QED hard-scattering corrections is
different from that of QCD. We expect that those divergences could be
canceled by soft photon radiative corrections. However, nobody knows how 
to include these radiative corrections in exclusive decay channels. As 
$\alpha_{em}$ is very small, we will neglect QED hard-scattering 
corrections in our computation.  
On the other hand, the penguin corrections are calculable 
not only for the case of QCD but also for that of QED. In quark level, 
considering only the contributions from tree operators, the penguin
corrections can be written in NDR scheme as\cite{fleischer}
\begin{eqnarray}
(\hat{m}_s(\mu))_{13}=(\hat{m}_s(\mu))_{15}=   
\frac{1}{2N}[-\frac{2}{3} + G(m_q, q, \mu)], \\
(\hat{m}_s(\mu))_{14}=(\hat{m}_s(\mu))_{16}=
-\frac{1}{2}[-\frac{2}{3} + G(m_q, q, \mu)], \\
(\hat{m}_e(\mu))_{27}=(\hat{m}_e(\mu))_{29}=
-\frac{4}{3}[-\frac{2}{3} + G(m_q, q, \mu)], \\
(\hat{m}_e(\mu))_{17}=(\hat{m}_e(\mu))_{19}=
-\frac{4}{9}[-\frac{2}{3} + G(m_q, q, \mu)],
\end{eqnarray}
with
\begin{equation}
G(m_q,q,\mu)=-4\int \limits_{0}^{1} dx ~x(1-x)
\ln \frac{m_q^2-x(1-x)q^2-i \epsilon}{\mu^2}.
\end{equation}
Here, $q^2$ in $G(m_q, q, \mu)$ remains uncertain for the
calculation of exclusive B decays. However, if we take
$q^2$ around $\frac{m_b^2}{2}$ in computations with conventional
factorization scheme, this variation does not change the results too
much\cite{Ali}. With the frame of QCD factorization in the heavy quark
limit, there will be no uncertainty for $q^2$, this will be shown below.
So, similar to what done in Ref.\cite{Beneke}, we will take both QCD and
QED penguin corrections into account. Surely, for the importance of
electroweak penguins in the decays $B\rightarrow \pi K$, we must take QED
penguin corrections into account as well.

Then in heavy quark limit, the amplitude for the decay of B to two light
pseudoscalar mesons $P$ and $P^{\prime}$ can be written as follows:
\begin{equation}
A(B\rightarrow P P^{\prime})=\frac{G_F}{\sqrt{2}}
\sum \limits_{p=u,c} \sum \limits_{i=1,10} v_p a^p_i 
\langle P P^{\prime} \vert Q_i \vert B \rangle_F,
\end{equation}
where $v_p$ is CKM factor and
$\langle P P^{\prime} \vert Q_i \vert B \rangle_F$ is the factorized
matrix elements which can be expressed by the product of the relevant
decay constant and form factor. Taking $a^c_1=a^c_2=0$ and assuming the
distribution amplitudes of light pseudoscalar mesons symmetric, we obtain 
the QCD
coefficients $a^p_i$ at next-to-leading order (NLO) in naive dimension
regularization (NDR) scheme (except $a_6$ and $a_8$ which read at leading 
order here for some special reasons). In order to cancel the scheme 
dependence in our calculation, we must take Wilson coefficients $C_i$ in 
NDR scheme as well. Then the explicit formulas of $a^p_i$ can be written
as
\begin{eqnarray}
a_1^u&=&C_1+\frac{C_2}{N} + \frac{\alpha_s}{4\pi} \frac{C_F}{N} C_2 F, \\
a_2^u&=&C_2+\frac{C_1}{N} + \frac{\alpha_s}{4\pi} \frac{C_F}{N} C_1 F,\\
a_3&=&C_3+\frac{C_4}{N} + \frac{\alpha_s}{4\pi} \frac{C_F}{N} C_4 F, \\
a_4^p&=&C_4+\frac{C_3}{N} -\frac{\alpha_s}{4\pi} \frac{C_F}{N} 
[(\frac{4}{3} C_1 +\frac{44}{3} C_3 + \frac{4f}{3} (C_4+C_6))
\ln \frac{\mu}{m_b}  
\nonumber \\
&&+ (G_P(s_p)-\frac{2}{3})C_1 
+(G_P(0)+G_P(1)-f_P^{I} - f_P^{II}+\frac{50}{3})C_3 
- \frac{2f}{3} C_4 \nonumber \\
&&+(3G_P(0)+G_P(s_c)+G_P(1))(C_4+C_6)+G_{P,8} C_{8G}], 
\\
a_5&=&C_5+\frac{C_6}{N}+\frac{\alpha_s}{4\pi}\frac{C_F}{N} C_6(-F-12),
\\
a_6&=&C_6+\frac{C_5}{N},\\
a_7&=&C_7+\frac{C_8}{N}+\frac{\alpha_s}{4\pi}\frac{C_F}{N} C_8(-F-12),
\\
a_8&=&C_8+\frac{C_7}{N}, \\
a_9&=&C_9+\frac{C_{10}}{N}+\frac{\alpha_s}{4\pi} \frac{C_F}{N} C_{10} F, \\
a_{10}^{p}&=&C_{10}+\frac{C_9}{N}+ 
\frac{\alpha_s}{4\pi} \frac{C_F}{N}C_{9} F \nonumber \\
&&+\frac{\alpha_{em}}{9\pi} [(-\frac{2}{3} (2(C_2+\frac{C_1}{N})+
(C_3+\frac{C_4}{N})+(C_5+\frac{C_6}{N}))\ln\frac{\mu}{m_b} \nonumber \\
&&+(\frac{2}{3}-G_P(s_p))(C_2+\frac{C_1}{N})
+(\frac{1}{3}-G_P(s_c)+\frac{G_P(1)}{2})(C_3+\frac{C_4}{N}) \nonumber \\
&&+(-G_P(s_c)+\frac{G_P(1)}{2})(C_5+\frac{C_6}{N})
-\frac{1}{2}C_{7\gamma} G_{P,8}].
\end{eqnarray}
Here $N=3$ ($f=5$) is the number of color (flavor), and 
$C_F=\frac{N^2-1}{2N}$ is the factor of color. We define the symbols in
the above expressions as the same as Beneke's, which are
\begin{eqnarray}
&&F=-12 \ln \frac{\mu}{m_b} -18+f_P^{I}+f_P^{II}, \\
&&f_P^{I}=\int \limits_{0}^{1}~ dx~g(x)\Phi_P(x), 
~G_{P,8}=\int \limits_{0}^{1}~ dx~G_8(x) \Phi_P(x), \\
&&G_P(s)=\int \limits_{0}^{1}~ dx~G(s,x) \Phi_P(x),
\end{eqnarray}
with 
\begin{eqnarray}
&&g(x)=3 \frac{1-2x}{1-x} \ln x - 3 i \pi, ~~G_8(x)=\frac{2}{1-x}, \\
&&G(s,x)=-4 \int \limits_{0}^{1}~ du~u(1-u) \ln (s-u(1-u)(1-x)-i 
\epsilon). 
\end{eqnarray}
Here $\Phi_{P}(x)$ is the leading twist distribution amplitude of the 
light meson, and $s_i=m_i^2/m_b^2$. ($m_i$ is the mass of quark appearing
in the penguin loop.) The contribution from the hard spectator scattering
are reduced into the factor $f_P^{II}$ which is written as

\begin{equation}
f_P^{II}=\frac{4 \pi^2}{N} 
\frac {f_{P^{\prime}}f_B}{F^{B\rightarrow P^{\prime}}_{+}(0) m_B^2}
\int \limits_{0}^{1}~ d\xi \frac{\Phi_B(\xi)}{\xi}
\int \limits_{0}^{1}~ dx~ \int \limits_{0}^{1}~ dy~ 
\frac{\Phi_P(x) \Phi_{P^{\prime}}(y) } {xy}.
\end{equation}
In above expression, $f_{P^{\prime}}$ ($f_B$) is the pseudoscalar meson 
(B meson) decay constant, $m_B$ the B meson mass, $F^{B\rightarrow 
P^{\prime}}_{+}(0)$ the $B\rightarrow P^{\prime}$ form factor at zero 
momentum transfer, and $\xi$ the light-cone momentum fraction of the 
spectator in the B meson.

One can find that our expressions are a little bit different from those in 
Ref.\cite{Beneke}. 
We think that there is an extra term $-\frac{2f}{3}C_4$ in Eq.(18)
which is missed in Eq.(8)
of Ref.\cite{Beneke}, but this difference is not large enough to make the 
numerical results  change too much. 
In addition, in the modes 
$B \rightarrow \pi K$, $f_P^{II}$ should be changed with the corresponding
meson which contains the spectator quark. But numerically the ratio 
$\frac{f_{\pi}}{F_{+}^{B\rightarrow \pi}}$ is nearly equal to
$\frac{f_{K}}{F_{+}^{B\rightarrow K}}$, so $f_{K}^{II}\simeq
f_{\pi}^{II}$, we will
not distinguish them in computation below. 

   For the coefficients $a_6$ and $a_8$, we want to give some comments.
One can see that the singularity of the hard-scattering function $G_8 (x)$
is at the endpoint $x=1$. If the distribution amplitudes did not 
fall off fast enough at the endpoints to suppress
these singularities in hard-scattering functions, our calculations would
not be infrared safe. For the case of $a_{1-5}$ and $a_{7,9,10}$, we
might use the twist-2 distribution amplitudes for pion or kaon which can 
cancel the
infrared divergences. Unfortunately, for the case of $a_{6,8}$, we must
employ the twist-3 distribution amplitudes. As pointed in 
Ref.\cite{Beneke,t3}, the
twist-3 distribution amplitudes for pion and kaon do not fall off at endpoints. 
So if this was true, the infrared divergences could not cancel in the
calculation. Therefore, as mentioned at the end of Ref.
\cite{Beneke}, the factorization formula breaks down in this case. 
We also noticed that in recent papers\cite{lihn,lucd},
the authors employed the twist-3 distribution amplitudes for pion and 
kaon which fall off fast enough at the endpoints, but that is just a 
model and not 
a prediction from  QCD sum rule or other non-perturbative approaches. So we 
still employ the asymptotic twist-3 distribution amplitudes of pion and kaon as 
$\Phi_P^3(x)=1$, then the singularities at the endpoints in 
hard-scattering kernels can not be suppressed by the distribution amplitudes 
in the case of $a_6$ and $a_8$. Like what Beneke {\it et al.} do,
we take $a_6$ and $a_8$ at leading order here. 

In the B rest frame, the two body decay width is
\begin{equation}  
\Gamma(B\rightarrow PP')=\frac {1}{8\pi}
\vert A(B\rightarrow P P') \vert ^2
\frac{\vert p \vert}{m_B^2},
\end{equation}
where
\begin{equation}
\vert p \vert =\frac
{\left[(m_B^2-(m_P+m_{P'})^2)(m_B^2-(m_P-m_{P'})^2)\right]^{\frac {1}{2}}}   
{2m_B}
\end{equation}    
is the magnitude of the momentum of the particle $P$ or $P'$. The
corresponding branching ratio is given by
\begin{equation}
BR(B\rightarrow PP')
=\frac{\Gamma(B\rightarrow PP')}{\Gamma_{tot}}.
\end{equation}  
The direct CP asymmetry ${\cal{A}}_{CP}$ for $B$ meson decays into $PP^{'}$
is defined as
\begin{equation}  
{\cal{A}}_{CP}=\frac {\vert A(B\rightarrow PP')	\vert ^2
-\vert A(\bar{B}\rightarrow \bar{P}\bar{P}')\vert ^2 }
{\vert A(B\rightarrow PP')\vert ^2
+\vert A(\bar{B}\rightarrow \bar{P}\bar{P}')\vert ^2 }~.
\end{equation}
Here we do not consider $B^0-\bar{B}^0$ mixing just for simplification.

    Because the momentum fraction distribution of the spectator quark
in B meson is peaked at $\Lambda_{QCD}/m_B$, we will take the distribution
amplitude for B meson as
\begin{equation}
\Phi_B(\xi)=\delta(\xi-\epsilon_B).
\end{equation}
$\frac{1}{\epsilon_B}$ is equal to the parameter $m_B/\lambda_B$ in
Ref.\cite{Beneke}. Here we take $\epsilon_B=0.05$. For simplification, we will
take the asymptotic form for the leading twist distribution amplitudes 
of pion and kaon as same as Ref\cite{Beneke}:
\begin{equation}
\Phi_{\pi}(x)=6x(1-x), ~~~~\Phi_{K}(y)=6y(1-y).
\end{equation}

After straightforward calculations,
we carry out the branching ratios and direct CP
asymmetries for the decays $B\rightarrow \pi \pi$ and $\pi K$ at two
different renormalization scales $\mu=5.0 ~GeV$ and $2.5~GeV$ which are
listed in Tab.2 and Tab.3 respectly. (Wilson coefficients in NDR scheme
\cite{Ali,ymz,buras} at two scales are listed in Tab.1.)
As a comparison, we also show the results with the
conventional factorization (BSW factorization and $N_{eff}=2$) in the 
last two tables. In our computation, we take the Wolfenstein parameters
of CKM matrix as follows: $A=0.8$, $\lambda=0.22$, $\rho=-0.12$,
$\eta=0.34$. The corresponding decay constants and form factors are taken
as follows\cite{Ali}: $f_{\pi}=0.13$ GeV, $f_K=0.16$ GeV;
$F^{B\rightarrow\pi}(0)=0.33$ and $F^{B \rightarrow K}(0)=0.38$.
Quark masses are taken as: $m_b=4.8~GeV$, $m_c=1.4~GeV$ and $m_s=105~MeV$.

From both Tab.2 and Tab.3, we find that QCD factorization enhance
the contributions from penguins so much that the branching ratios of
$B\rightarrow \pi K$ become larger nearly by a factor of 2 than the
estimate of the generalized factorization. It is because that the QCD
and QED corrections are constructive to the amplitudes with the generalized
factorization and they enhance the branching ratios. One also see that
the contributions from electroweak penguins can not be neglected in the decays
$B_u^- \rightarrow \pi^0 K^-$ and $\bar{B}_d^0 \rightarrow \pi^0 \bar{K}^0$
no matter with the generalized factorization or QCD factorization,
because the amplitudes of these two modes contain a term of the coefficient
$a_9$ which is large enough to compare with the coefficients of QCD penguins.
For the case of $\bar{B}_d^0 \rightarrow \pi^{+} \pi^{-}$,
the differences between these two factorization
schemes are not too apparent because that this decay mode is
dominated by the coefficient $a_1$. The radiative correction to $a_1$ at the 
order of $\alpha_s$ is very small comparing with the leading order part 
of $a_1$. This is similar to that the effective coefficient 
$a^{eff}_1=C_1+C_2/N_{eff}$ in generalized factorization is unsensitive
to the variation of the phenomenological parameter $N_{eff}$. But for the 
coefficient $a_2$, it is very different for its small leading order 
part. So for the modes $B_u^- \rightarrow \pi^0 \pi^-$ and 
$\bar{B}_d^0 \rightarrow \pi^0 \pi^0$, the differences between these two
factorization schemes are very large.

For the case of ${\cal{A}}_{CP}$, the differences between generalized 
factorization and QCD factorization in the heavy quark limit are quite 
large because in QCD factorization the imaginary parts enter the decay 
amplitudes through hard scattering kernels contribution. Then the strong 
phases of some $a_i$ (i=even number) dominant
processes may be changed very much, then CP asymmetries of these modes
can be dramatically large. One can see that CP asymmetries of
$\bar{B}_d^0 \rightarrow \pi^0 \pi^0$ is about 80\%, 
but this magnificent direct CP asymmetry is very hard to observe for
its small branching ratio. For the modes $B \rightarrow \pi K$, 
the CP asymmetries change much as well. In the modes of 
$B \rightarrow\pi K$, the signs and magnitudes of
CP asymmetries are changed comparing with the results of the generalized
factorization.

  The authors of Ref.\cite{Beneke} pointed out that the amplitudes
derived from the QCD factorization in the heavy quark limit are
independent of the renormalization scale physically. Numerically, we still
find that the dependences in our results of the branching ratios for the
decays $B \rightarrow \pi K$ are visible. Comparing the results at the
scale $\mu=5.0~GeV$ and $\mu=2.5 ~GeV$, this variation brings about 
$\pm 20\%$ uncertainty to the estimates for the branching ratios of 
$B \rightarrow \pi K$. As shown in Ref.\cite{Beneke} and
our paper, the scale dependences of the results $B \rightarrow \pi \pi$ are
small. In recent calculation of $\bar{B} \rightarrow D^{(\star)} \pi^-$ in
the heavy quark limit\cite{korea}, the scale dependences are also small.
Maybe for the case of pure tree or tree dominant processes, the computation
with the frame of the QCD factorization cancels the dependence of the
renormalization scale very well. But for some pure penguin or penguin
dominant processes, the scale dependences are visible. However, these
dependences are smaller than the estimates with the generalized
factorization. 

We also show the dependences of the branching ratios and  
direct CP asymmetries on the weak phase 
$\gamma=\arg V_{ub}^{\star}$ respectly in Fig.2 and Fig.3. 
In both Fig.2 and Fig.3, the results of (a) and (b) are 
carried out with the QCD factorization in the heavy
quark limit at the scale $\mu=2.5~GeV$. 

From Fig.2 and Fig3, we find that the results are in favor of the
experimental measurements when $90^{\circ}<\gamma<270^{\circ}$ because the
branching ratio of $\bar{B}_d^0 \rightarrow \pi^+ \pi^-$ becomes small and
the branching ratios of
$B_u^- \rightarrow \pi^{0} K^{-}$,
$B_u^{-} \rightarrow \pi^{-} \bar{K}^{0}$ and
$\bar{B}_d^0 \rightarrow \pi^{+} K^{-}$
are closer in that region. This is consistent with the fit for $\gamma$ of 
CLEO and other researchers\cite{CLEO,hxg}. 

We note that our estimate of branching ratio of $\bar{B}_d^0 \rightarrow 
\pi^+ \pi^-$ seems larger than the experimental measurement even if we take
$\gamma>90^{\circ}$. And the branching ratio of
$\bar{B}_d^0 \rightarrow \pi^0 \bar{K}^0$ is about 3 times smaller than the
central value of experimental result and unsensitive to the variation of
$\gamma$. Due to uncertainties
of the form factors $F_{+}^{B\rightarrow \pi}$ and $F_{+}^{B\rightarrow K}$,
we try to vary these form factors in a relevant region to make the
branching
ratio $\bar{B}_d^0 \rightarrow \pi^+ \pi^-$ smaller while that of 
$\bar{B}_d^0 \rightarrow \pi^0 \bar{K}^0$ is larger.
We find the branching ratios of $B\rightarrow \pi \pi$ and
$\pi K$ are rather sensitive to the form factor $F_{+}^{B\rightarrow \pi}$
than $F_{+}^{B\rightarrow K}$. And both of them increase with 
$F_{+}^{B\rightarrow \pi}$. So our attempt faces a failure. 
But now the errors in present measurements of CLEO are still large, and
some uncertainties remain in theoretical frame, such as, the light-cone
distribution amplitudes of the mesons, the heavy to light transition form 
factors and etc. Therefore, the disagreement between our prediction and
the present experimental measurement is not so significant. It needs us
more detailed study with the improved experimental measurements and theoretical 
approaches in future.

   In conclusion, QCD factorization can give an estimate of 
strong phases in B charmless decays from final state hard scattering.
This might be beneficial to extracting the weak phases from CP asymmetries 
in B decays. But due to the theoretical uncertainties such as meson
light-cone distribution amplitudes and heavy to light transition form factors, 
the prediction for branching ratios within the frame of QCD factorization  
remains a little bit ambiguous. On the other hand, since that $m_b$ is not
a very large scale, maybe the corrections at the order of 
$\Lambda_{QCD}/m_b$ are needed. So, 
how to develop a complete perturbative frame of heavy quark expansion in 
heavy to light decays will be a potentially interesting and beneficial work.

\section*{Acknowledgement}
We thank Dr. Z.T. Wei for helpful discussions.
   
\narrowtext
\tighten


\begin{table}
\vspace*{1cm}
\begin{tabular}{c|cccccc}
$~$ & $C_1$ & $C_2$ & $C_3$ & $C_4$ & $C_5$ & $C_6$ \\ \hline
$\mu=5.0 ~GeV$ & $1.150$ & $-0.313$ & $0.016$ & $-0.033$ & $0.009$ & 
$-0.042$ \\ \hline 
$\mu=2.5 ~GeV$ & $1.117$ & $-0.257$ & $0.017$ & $-0.044$ & $0.011$ & 
$-0.056$ 
\end{tabular}

\vspace{0.5cm}
\begin{tabular}{c|cccccc}
$~$ & $C_7$ & $C_8$ & $C_9$ & $C_{10}$ & $C_{7\gamma}$ & $C_{8G}$ \\ \hline
$\mu=5.0 ~GeV$ & $-2 \times 10^{-5}$ & $38 \times 10^{-5}$ & $-0.0103$ & 
$0.0021$ & $-0.300$ & $-0.144$ \\ \hline 
$\mu=2.5 ~GeV$ & $-1 \times 10^{-5}$ & $50\times 10^{-5}$ & $-0.010$ 
& $0.002$ & $-0.336$ & $-0.158$ 
\end{tabular}
\vspace{0.5cm}

\caption{Wilson coefficients in NDR scheme}

\end{table}

\begin{table}

\vspace*{1cm}

\begin{tabular}{l|cccc}
$\mu=5.0~GeV$ &\multicolumn{2}{c}{Branching~ Ratios} &
\multicolumn{2}{c}{CP~Asymmetries} \\
Decay Modes & Generalized FA & QCD FA& Generalized
FA & QCD FA\\\hline
$B_u^{-}\rightarrow \pi^0 \pi^{-}$ & $6.33(6.71)$ & $4.30(4.60)$ &
$-0.1(0)$ & $0(0)$ \\
$\bar{B}^0_d \rightarrow \pi^{+} \pi^{-}$ & $7.36(7.45)$ & $7.93(7.97)$ &
$2.4(2.5)$ & $-3.2(-3.6)$ \\
$\bar{B}^0_d \rightarrow \pi^0 \pi^0$ & $0.38(0.49)$ & $0.14(0.21)$ &
$-7.5(-5.7)$ & $78.4(64.7)$ \\
\hline
$B_u^{-} \rightarrow \pi^0 K^{-}$& $6.00(3.81)$ & $10.2(7.10)$ &
$-4.5(-7.1)$ & $4.1(5.7)$ \\
$B_u^{-} \rightarrow \pi^{-} \bar{K}^0$& $5.50(5.81)$ & $12.3(11.1)$ &
$-0.1(-0.1)$ & $0.4(0.4)$
\\
$\bar{B}^0_d \rightarrow \pi^{+} K^{-}$ & $7.80(7.01)$ & $15.0(13.9)$ &
$-5.4(-6.1)$ & $2.4(3.0)$\\
$\bar{B}^0_d \rightarrow \pi^0 \bar{K}^0$ & $1.47(2.7)$ & $3.90(5.20)$ &
$3.8(4.0)$ & $-3.2(-2.8)$
\end{tabular}
\vspace*{0.5cm}
\caption{In this table, branching ratios are in the unit of $10^{-6}$ and CP
asymmetries are in the unit of one percent. 
And the values in the brackets are the results without considering the
contributions of EW penguins.}

\end{table}


\begin{table}

\begin{tabular}{l|cccc}
$\mu=2.5~GeV$ &\multicolumn{2}{c}{Branching~ Ratios} &
\multicolumn{2}{c}{CP~Asymmetries} \\
Decay Modes & Generalized FA & QCD FA & Generalized
FA &QCD FA \\\hline
$B_u^{-}\rightarrow \pi^0 \pi^{-}$ & $6.07(7.08)$ & $4.41(4.65)$ &
$-0.1(0)$ & $0(0)$ \\
$\bar{B}^0_d \rightarrow \pi^{+} \pi^{-}$ & $6.94(7.02)$ & $7.54(7.50)$ &
$2.8(2.8)$ & $-3.2(-3.8)$ \\
$\bar{B}^0_d \rightarrow \pi^0 \pi^0$ & $0.63(0.77)$ & $0.26(0.37)$ &
$-5.4(-4.4)$ & $73.4(61.1)$ \\
\hline
$B_u^{-} \rightarrow \pi^0 K^{-}$& $8.50(5.89)$ & $11.9(9.30)$ &
$-3.2(-4.7)$ & $4.3(5.3)$ \\
$B_u^{-} \rightarrow \pi^{-} \bar{K}^0$& $9.08(9.44)$ & $15.2(15.1)$ &
$-0.1(-0.1)$ & $0.4(0.4)$
\\
$\bar{B}^0_d \rightarrow \pi^{+} K^{-}$ & $11.90(11.07)$ & $17.9(18.1)$ &
$-3.5(-3.9)$ & $1.9(2.2)$\\
$\bar{B}^0_d \rightarrow \pi^0 \bar{K}^0$ & $2.78(4.41)$ & $5.00(7.00)$ &
$2.2(1.3)$ & $-3.8(-3.2)$\
\end{tabular}
\vspace*{0.5cm}
\caption{In this table, branching ratios are in the unit of $10^{-6}$ and CP
asymmetries are in the unit of one percent. 
And the values in the brackets are the results without considering the
contributions of EW penguins.}

\end{table}


\begin{figure}[tb]
\vspace*{1cm}
\centerline{\epsfig{figure=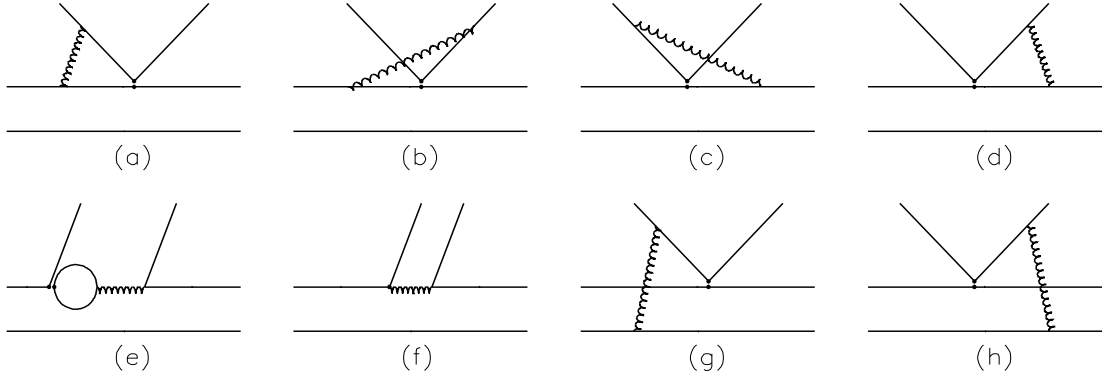,height=6cm,width=15cm,angle=0}}
\vspace*{1.cm}
\caption{Order of $\alpha_s$ corrections to hard-scattering kernels $T^I$
and $T^{II}$. The upward quark lines represent the ejected quark pairs
from b quark weak decays.}
\end{figure}

\begin{figure}[htb]
\vspace*{2cm}
\centerline{\epsfig{figure=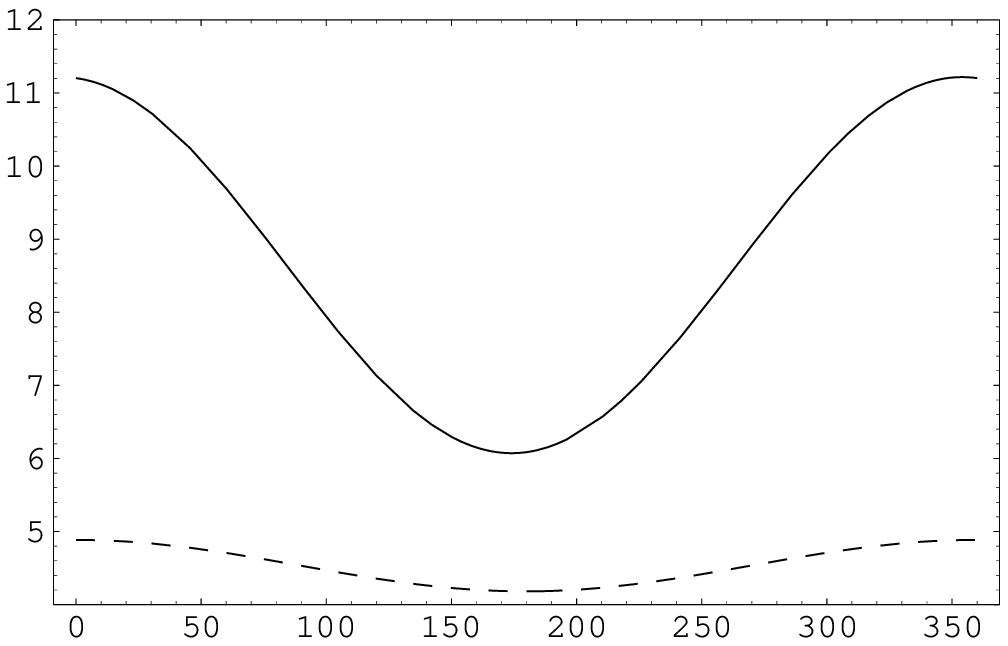,height=5cm,width=8cm,angle=0}
            \epsfig{figure=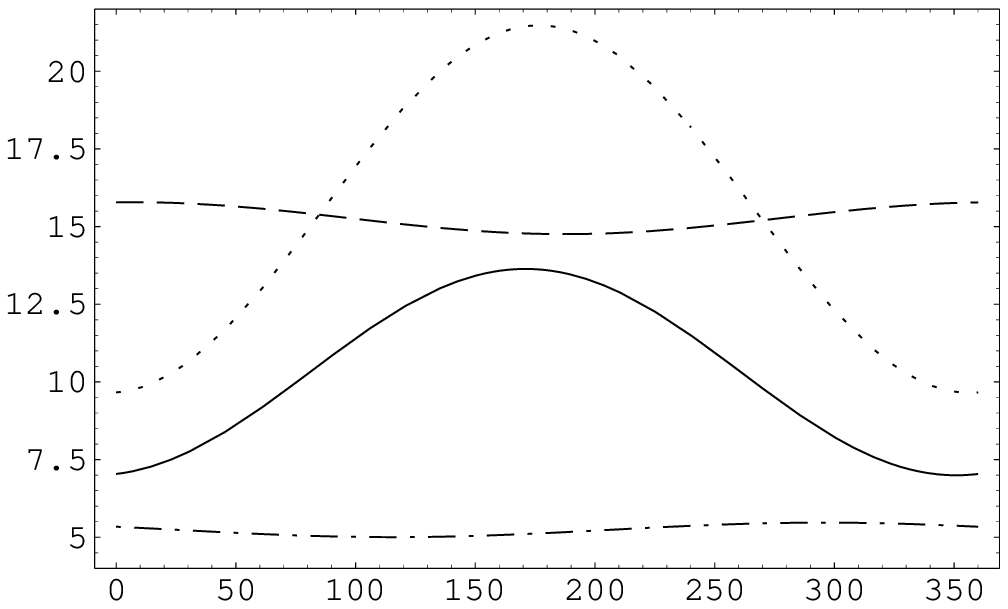,height=5cm,width=8cm,angle=0}}
\centerline{(a)~~~~~~~~~~~~~~~~~~~~~~~~~~~~~~~~~~~~~~~~~~~~~~~~~~~~~(b)}
\vspace*{1cm}
\caption{
$Br(B \rightarrow \pi \pi) \times 10^{6}$ vs.
$\gamma=\arg{V^{\star}_{ub}}$ (Fig1.(a)) and
$Br(B \rightarrow \pi K) \times 10^6$ vs. $\gamma$ (Fig.(b)). 
In Fig.(a), solid, dashed curves are for  
$\bar{B}_d^0 \rightarrow \pi^+ \pi^-$ and $B_u^-\rightarrow \pi^0 \pi^-$
respectly;
in Fig.(b), solid, dashed, dotted and
dot-dashed curves are for $B_u^{-} \rightarrow \pi^{0} K^{-}$, 
$\pi^{-} \bar{K}^0$ and $\bar{B}_d^0 \rightarrow \pi^{+} K^{-}$,
$\pi^0 \bar{K}^0$ respectly.}

\end{figure}


\begin{figure}[htb]
\vspace*{2cm}
\centerline{\epsfig{figure=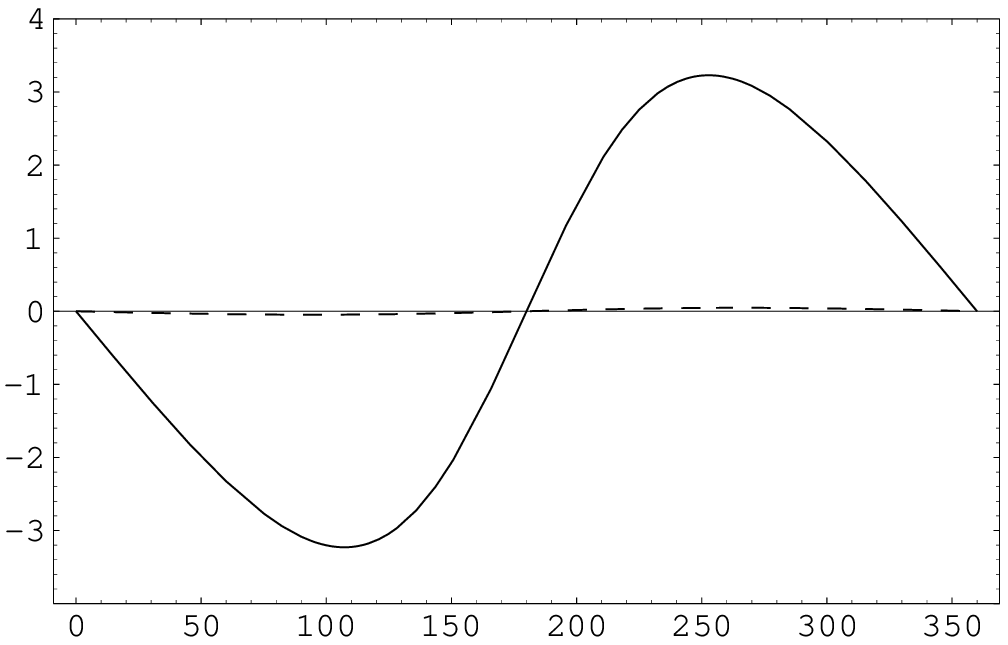,height=5cm,width=8cm,angle=0}
            \epsfig{figure=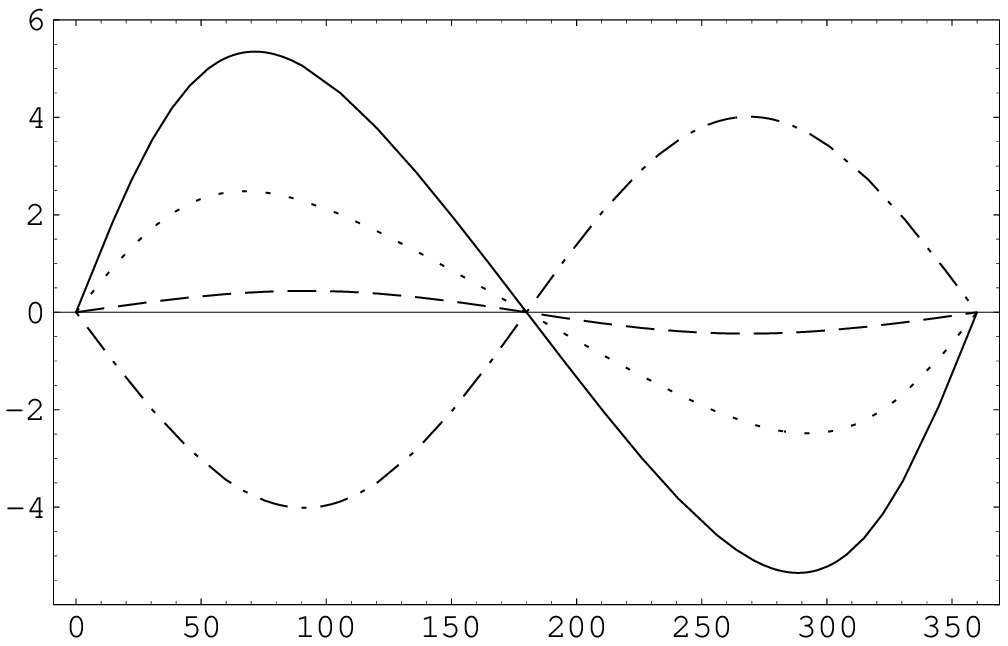,height=5cm,width=8cm,angle=0}}
\centerline{(a)~~~~~~~~~~~~~~~~~~~~~~~~~~~~~~~~~~~~~~~~~~~~~~~~~~~~~(b)}
\vspace*{1cm} 
\caption{$A_{CP}(B \rightarrow \pi \pi) \times 10^{2}$ vs.
$\gamma=\arg{V^{\star}_{ub}}$ (Fig1.(a)) and $A_{CP}(B \rightarrow \pi K)
\times 10^2$ vs. $\gamma$ (Fig.(b)).  
In Fig.(a), solid, dashed curves are
for $\bar{B}_d^0 \rightarrow \pi^+ \pi^-$ and $B_u^-\rightarrow \pi^0
\pi^-$ respectly;
in Fig.(b), solid, dashed, dotted and dot-dashed curves are for
$B_u^{-} \rightarrow \pi^{0} K^{-}$, $\pi^{-} \bar{K}^0$ and $\bar{B}_d^0
\rightarrow \pi^{+} K^{-}$, $\pi^0 \bar{K}^0$ respectly.} 
\end{figure}

\end{document}